\documentclass[nopreprintline]{elsarticle}
\usepackage[utf8]{inputenc}
\usepackage{hyperref,amsmath}
\usepackage{tablefootnote}
\usepackage{lmodern} % font improvements

\usepackage{mmacells} % Mathematica code
\mmaSet{moredefined={MultivariateApart,MultivariateAbbreviatedApart,AbbreviateDenominators,ApartOrder,ApartBasis,ApartReduce,ApartTemporaryDirectory,WriteSingularBasisInput,WriteFormProcedure}}
\mmaDefineMathReplacement[≤]{<=}{\leq}
\mmaDefineMathReplacement[≥]{>=}{\geq}
\mmaDefineMathReplacement[≠]{!=}{\neq}
\mmaDefineMathReplacement[→]{->}{\to}[2]
\mmaDefineMathReplacement[⧴]{:>}{:\hspace{-.2em}\to}[2]
\mmaDefineMathReplacement{∉}{\notin}
\mmaDefineMathReplacement{∞}{\infty}
\mmaDefineMathReplacement{��}{\mathbbm{d}}
\newcommand{\textmma}[1]{\mmaInlineCellNonVerb{Code}{#1}}

\usepackage{textcomp} % centered tilde
\newcommand{\textapprox}{\raisebox{0.5ex}{\texttildelow}}

\newdefinition{alg}{Algorithm}
\newdefinition{recipe}{Recipe}
\newdefinition{definition}{Definition}
\newcommand{\Leinartas}{Le{\u\i}nartas}
\newcommand{\mathematica}{\texttt{Mathematica}}
\newcommand{\form}{\texttt{Form}}
\newcommand{\singular}{\texttt{Singular}}
\newcommand{\lt}[0]{{\text{LT}}}

\definecolor{darkred}{rgb}{0.9,0,0}
\definecolor{darkblue}{rgb}{0,0,0.9}

\journal{Computer Physics Communications}

\begin{document}
\allowdisplaybreaks[3]

\begin{frontmatter}

\title{\texorpdfstring{\hfill{\normalsize MITP/21-002, MSUHEP-20-016}\\[7ex]}{}
MultivariateApart: Generalized Partial Fractions}

\author[1]{Matthias Heller}
\ead{maheller@students.uni-mainz.de}

\author[2]{Andreas von Manteuffel}
\ead{vmante@msu.edu}

\address[1]{PRISMA$^+$ Cluster of Excellence, Johannes-Gutenberg Universit\"{a}t,\\
55099 Mainz, Deutschland}
\address[2]{Department of Physics and Astronomy, Michigan State University,\\
East Lansing, Michigan 48824, USA}

\begin{abstract}
We present a package to perform partial fraction decompositions of multivariate rational functions.
The algorithm allows to systematically avoid spurious denominator factors
and is capable of producing unique results also when being applied to terms of a sum separately.
The package is designed to work in Mathematica, but also provides interfaces to the Form and Singular computer algebra systems.
\end{abstract}
\end{frontmatter}

\newpage
\tableofcontents

%%%%%%%%%%%%%%%%%%%%%%%%%%%%%%%%%%%%%%%%%%%%%%%%%%%%%%%%%%%%%%%%%%%%%%%%%%%%%%%%%%%%%%%%
\section{Introduction}
\label{sec:intro}
%%%%%%%%%%%%%%%%%%%%%%%%%%%%%%%%%%%%%%%%%%%%%%%%%%%%%%%%%%%%%%%%%%%%%%%%%%%%%%%%%%%%%%%%

Employing partial fraction decompositions to bring analytic expressions into a unique form has a long history in particle physics. Already one of the first computer algebra systems (CAS), Veltman's program \texttt{Schoonschip}~\cite{Strubbe:1974vj}, that was developed in 1963 to calculate radiative corrections, had a function to partial fraction products of two or three rational functions~\cite{RemiddiTini80Fest}.
With \texttt{Schoonship}'s successor {\form} \cite{Vermaseren:2000nd,Ruijl:2017dtg}, partial fraction decompositions of rational functions became widely established in the particle physics community.
While the standard partial fraction decomposition is a method for rational functions of a single variable, it is often applied iteratively to treat multivariate functions.
However, this generalization has two drawbacks: first this method generally introduces new denominator factors, i.e.\ spurious singularities, and second, it is often slower than reduction schemes in which spurious singularities are avoided from the beginning.
An alternative representation which avoids spurious denominators is known as {\Leinartas}' decomposition \cite{Leinartas:1978pf,Raichev:2012pf,Meyer:2016slj,Meyer:2017joq}.
It separates denominators which don't share common zeros or are algebraically dependent.

Here, we discuss a method to compute a partial fraction decomposition via polynomial reductions.
Our general approach coincides with that of Ref.\ \cite{Abreu:2019odu}, but we present new insights concerning the choice of a good monomial ordering and the resulting output forms.
In particular, we propose a new monomial ordering, which still guarantees the separation of denominator zeros but allows for deviations from {\Leinartas}' form to allow for lower denominator degrees.
We also discuss performance optimizations and other practical issues, such as the local elimination of specific denominators, if they are known to be spurious globally.
We implement our algorithms in a {\mathematica} package which is publicly available. It can be used to partial fraction rational functions directly in {\mathematica}, but also allows to generate local replacement rules that can be incorporated in {\form}.

The outline of the paper is as follows. In Section\ \ref{sec:motivation} we motivate our method.
We demonstrate the problem with iterated univariate partial fractions and discuss ambiguities when applying the ``classical'' {\Leinartas} decomposition to individual terms of a sum.
We spell out an explicit wish-list for an ideal partial fraction algorithm and discuss which items are addressed by available implementations and the one described in this article, respectively.
In Section\ \ref{sec:alg} we introduce our algorithms in detail, give examples and show how it can be used efficiently in complicated cases.
Furthermore, we present some considerations concerning the reconstruction of rational functions from finite field samples, which can be useful to speed-up tasks like linear system solving.
In Section\ \ref{sec:package} we give an introduction to our {\mathematica} package. 
We describe the relevant functions and give examples for their usage.
In Section\ \ref{sec:applications} we list ``real world'' applications of our methods to the calculation of Feynman amplitudes, often resulting in significantly reduced sizes of the mathematical expressions.
In Section\ \ref{sec:outlook} we conclude and provide an outlook.
Appendix\ref{sec:polyred} gives a brief introduction to polynomial reductions.
Appendix\ref{sec:leinartaspolyred} discusses the relation between {\Leinartas}' decomposition and our polynomial reduction method.

%%%%%%%%%%%%%%%%%%%%%%%%%%%%%%%%%%%%%%%%%%%%%%%%%%%%%%%%%%%%%%%%%%%%%%%%%%%%%%%%%%%%%%%%
\section{Motivation}
\label{sec:motivation}
%%%%%%%%%%%%%%%%%%%%%%%%%%%%%%%%%%%%%%%%%%%%%%%%%%%%%%%%%%%%%%%%%%%%%%%%%%%%%%%%%%%%%%%%
\subsection{Spurious poles in iterated partial fractions\label{Sec:iterated_PF}}
%%%%%%%%%%%%%%%%%%%%%%%%%%%%%%%%%%%%%%%%%%%%%%%%%%%%%%%%%%%%%%%%%%%%%%%%%%%%%%%%%%%%%%%%

We consider a univariate rational function $r(x)=n(x)/d(x)$, where $n$ and $d$ are polynomials in $x$ and the coefficient field $K$ could be e.g.\ the set of rational numbers.
The denominator $d(x)$ can be factored into irreducible factors $d_i(x)$ such that $d(x)=\prod_i d_i^{\alpha_i}(x)$, where the polynomials $d_i$ can not be written as a product of two non-constant polynomials and $\alpha_i\in\mathbbm{N}$.
The \emph{univariate partial fraction decomposition} of $r(x)$ is given by
\begin{equation}
    r(x) = \sum_{i} \sum_{j\leq \alpha_i} \frac{n_i(x)}{d_i^j(x)},
\end{equation}
where in each term the degree of the numerator is smaller than the degree of the denominator. As an example, consider the partial fraction decomposition
\begin{equation}
   %r(x) =
   \frac{x}{(x - 1) (x + 1)^2} =-\frac{1}{4 (x + 1)} + \frac{1}{2 (x + 1)^2} + \frac{1}{4 (x - 1)}.
\end{equation}

In the case of multivariate rational functions $r(x,y,\ldots)$ the partial fractioning can be performed iteratively in each variable.
One performs a partial fraction decomposition with respect to the first variable $x$, treating the other variables as constants during this step.
The resulting coefficients $p_i$ are now rational functions of the remaining variables, and one can perform a partial fraction decomposition of the $p_i(y,\ldots)$ with respect to the next variable $y$.
One iterates until the coefficients are actual numbers.
As an example, consider the function $r(x,y) = {1/((x-f(y))(x-g(y)))}$, where $f(y)$ and $g(y)$ are two different polynomials of $y$.
A partial fractioning with respect to $x$ gives
\begin{equation}
   \frac{1}{(x-f(y))(x-g(y))} = \frac{1}{(f(y)-g(y))}\frac{1}{(x-f(y))} - \frac{1}{(f(y)-g(y))}\frac{1}{(x-g(y))}.
\label{Eq:iterativePF}
\end{equation}
The partial fractioning in $x$ thus introduces a denominator $f(y)-g(y)$, which may have zeros at regular points of the original rational function $r(x,y)$.
For example, consider the special case
\begin{equation}\label{eq:pf_iterated}
   \frac{1}{(x+y)(x-y)} = \frac{1}{2y}\frac{1}{(x-y)}-\frac{1}{2y}\frac{1}{(x+y)},
\end{equation}
where the subsequent partial fractioning of the $x$-independent coefficients in $y$ is trivial.
Although the original expression is manifestly regular at $y=0$, the individual terms of the partial fractioned form are not---they introduce spurious poles $1/y$.
This obscures the interpretation of the singularity structure of the expression and can lead to loss of precision in its numerical evaluation close to $y=0$.
Note that a different spurious pole would have appeared in this example if we had chosen to first partial fraction in $y$ and then in $x$.

We conclude that iterated partial fractioning will in general introduce spurious poles and is therefore not an ideal approach to the multivariate case.

%%%%%%%%%%%%%%%%%%%%%%%%%%%%%%%%%%%%%%%%%%%%%%%%%%%%%%%%%%%%%%%%%%%%%%%%%%%%%%%%%%%%%%%%
\subsection{Features of \texorpdfstring{{\Leinartas}'}{Leinartas'} decomposition}
\label{sec:leinartas}
%%%%%%%%%%%%%%%%%%%%%%%%%%%%%%%%%%%%%%%%%%%%%%%%%%%%%%%%%%%%%%%%%%%%%%%%%%%%%%%%%%%%%%%%

{\Leinartas}' decomposition is an approach to multivariate partial fractioning, which avoids the introduction of spurious singularities.
In this Section, we review this decomposition and comment on possible ambiguities arising in practical calculations.
Polynomial reductions and the usage of Gr\"obner bases will play an important role here and in the following; in case the reader is not familiar with these concepts we recommend to read Appendix\ref{sec:polyred} first.

\begin{definition}
\label{def:leinartas}
\emph{{\Leinartas}' decomposition} \cite{Leinartas:1978pf,Raichev:2012pf} of a rational 
function $r$ of the variables $x_1,\ldots,x_n$ with coefficients in the field $K$ is a decomposition of the form
\begin{equation}
\label{eq:leinartas}
    r(x_1,\ldots)=\sum_\mathcal{S}\frac{n_\mathcal{S}(x_1,\ldots)}{\prod_{i\in \mathcal{S}} d_i^{\alpha_i}(x_1,\ldots)},
\end{equation}
where, for each term of the decomposition, $\mathcal{S}$ is an index set, such that all denominators $d_i$ with $i\in\mathcal{S}$ 
\begin{itemize}
    \item[(i)] have common zeros in $\overline{K}^n$, and
    \item[(ii)] are algebraically independent.
\end{itemize}
Here, $\overline{K}$ denotes the algebraic closure of $K$, e.g.\ the algebraic numbers for $K=\mathbbm{Q}$ or $\mathbbm{C}$ for $K=\mathbbm{R}$.
\end{definition}

A set of polynomials $\{d_1,\ldots,d_m\}$ is called \emph{algebraically dependent} if there exists a non-zero polynomial $A$ in $m$ variables such that $A(d_1,\ldots,d_m)=0$, see e.g.\  \cite{CoxLittleOShea,KreuzerRobbianoBook1}.
$A$ is called the \emph{annihilator} of the ideal generated by the polynomials $\{d_1,\ldots,d_m\}$.
The annihilator can be obtained by calculating the Gr\"obner basis of the ideal $\{y_1-d_1(x_1,\ldots,x_n),\ldots, y_m-d_m(x_1,\ldots,x_n)\}$, if one chooses a monomial ordering in which $y_i \prec x_j\; \forall i,j$.
A set of polynomials is called \emph{algebraically independent} if it is not algebraically dependent.
Since for $n$ variables at most $n$ polynomials are algebraically independent, there can be at most $n$ different denominator factors for each term in \eqref{eq:leinartas}.

\begin{alg}
\label{alg:leinartas}
\emph{{\Leinartas}' decomposition} of a rational function can be reached in two reduction steps \cite{Raichev:2012pf}:
\begin{enumerate}
    \item Use Hilbert's Nullstellensatz to decompose the denominator of $r$ into several terms such that each term fulfills (i).
    \item Calculate the annihilator for each term and use it to decompose each denominator to reach (ii).
\end{enumerate}
\end{alg}

In contrast to iterated partial fractioning, these decomposition steps do not introduce new singularities.
Appendix\ref{sec:leinartaspolyred} gives more details for this algorithm.
A decomposition that fulfills requirements (i) and (ii) is not unique.
In order to resolve this ambiguity, Refs.\ \cite{Meyer:2017joq,Boehm:2020ijp} require additionally for each term \eqref{eq:leinartas}, that the numerator $p_\mathcal{S}$ is reduced with respect to the ideal generated by the denominators $\{d_i | i\in \mathcal{S} \}$.
We note that the algorithm of \cite{Boehm:2020ijp} assumes the input expression to be a single rational function $r=n/d$ with polynomials $n$ and $d$ and starts with a factorization of $d$.

Further ambiguities can arise due to spurious denominators which are not automatically removed.
The algorithm in \cite{Boehm:2020ijp} for example does not guarantee a unique result, if $n$ and $d$ are not coprime.
However, common factors between numerator and denominator can always be eliminated by first performing a greatest-common-divisor computation.
% demo of this issue in Singular's pfd:
% ring r = 0,(x,y,z),dp;
% poly g=(x-y)*(y+z)*(x+y+2*z);
% poly f=x+y+2*z;
% list d=pfd(f,g);
% gives 1 / (q2*q3) + 2 / (q1*q3) where q1 = x-y, q2 = y+z, q3 = x+y+2*z
% but f/g = 1/(q1*q2) doesn't need q3

In practical applications one also encounters sums of rational functions, and it can be useful to decompose individual terms of the sum separately instead of combining them over a common denominator first.
With the methods spelled out so far, uniqueness is not guaranteed in this approach, even in absence of any spurious denominator.
Let us illustrate this phenomenon by considering an example where
\begin{equation}
    r(x,y)=\frac{2x-y}{x(x+y)(x-y)}.
\label{eq:rexampleambig}
\end{equation}
Then a {\Leinartas} decomposition of $r$ is given by
\begin{equation}\label{eq:red1}
    r(x,y)=\frac{1}{x(x + y)} + \frac{1}{(x - y)(x + y)}.
\end{equation}
Another {\Leinartas} decomposition is given by
\begin{equation}\label{eq:red2}
    r(x,y)=\frac{3}{2x(x+y)}+\frac{1}{2x(x-y)}.
\end{equation}
If we consider the terms in Eqs.\ \eqref{eq:red1} and \eqref{eq:red2} separately, each of them is in {\Leinartas}'decomposed form and the numerator is reduced with respect to the denominator, but the resulting representation is different for the respective sums.
In order resolve this ambiguity, some kind of ``global'' information needs to be incorporated into the decomposition of ``local'' terms.

%%%%%%%%%%%%%%%%%%%%%%%%%%%%%%%%%%%%%%%%%%%%%%%%%%%%%%%%%%%%%%%%%%%%%%%%%%%%%%%%%%%%%%%%
\subsection{A wish-list for a ``good'' partial fractioning algorithm}
\label{sec:wishlist}
%%%%%%%%%%%%%%%%%%%%%%%%%%%%%%%%%%%%%%%%%%%%%%%%%%%%%%%%%%%%%%%%%%%%%%%%%%%%%%%%%%%%%%%%

Lets assume someone gives us a rational expression in any form, expanded, over a common denominator, or partially mixed.
Then an ``ideal'' partial fraction decomposition would have the following properties
\begin{itemize}
    \item[(i)] it should give a unique answer,
    \item[(ii)] it should not introduce spurious denominator factors,
    \item[(iii)] it should commute with summation,
    \item[(iv)] it should eliminate spurious denominators if they are present in the input.
\end{itemize}

\noindent
Note that requirement (iii) can be crucial, if one aims at employing such a partial fraction algorithm in a system like {\form}.
In such a case one usually starts with fully expanded expressions, uses local replacement rules, and then wants to obtain a unique answer, such that cancellations can take place.

Unfortunately, there is no know solution to fulfil all of these points simultaneously.
In the next Section, we consider an algorithm that fulfills requirements (i), (ii) and (iii), and, in cases in which one knows which denominators are spurious in the final answer, also point (iv).

%%%%%%%%%%%%%%%%%%%%%%%%%%%%%%%%%%%%%%%%%%%%%%%%%%%%%%%%%%%%%%%%%%%%%%%%%%%%%%%%%%%%%%%%
\section{Multivariate partial fractions with polynomial reduction\label{sec:alg}}
%%%%%%%%%%%%%%%%%%%%%%%%%%%%%%%%%%%%%%%%%%%%%%%%%%%%%%%%%%%%%%%%%%%%%%%%%%%%%%%%%%%%%%%%
\subsection{Reduction algorithm}
%%%%%%%%%%%%%%%%%%%%%%%%%%%%%%%%%%%%%%%%%%%%%%%%%%%%%%%%%%%%%%%%%%%%%%%%%%%%%%%%%%%%%%%%

Let $\{d_1,...d_m\}$ be the irreducible denominators of a rational function or a sum of rational functions with $d_i \in K[x_1,\ldots,x_n]$.
Consider the ideal
\begin{equation}
I = \left< q_1 d_1(x_1,\ldots)-1,\ldots,q_m d_m(x_1,\ldots)-1\right>
\label{Eq:ideal},
\end{equation}
where $I \subset K[q_1,\ldots,q_m,x_1,\ldots,x_n]$ and $q_i$ label inverse denominators.
The main idea here is that setting the generators of the ideal $q_i d_i(x_1,\ldots)-1$ to zero corresponds to the relation $q_i = 1/d_i(x_1,\ldots)$.
If we rewrite a rational function $r$ as a polynomial in the variables $q_1,\ldots,x_1,\ldots$ and reduce it with respect to the ideal $I$, we do not introduce new denominator factors and obtain a unique representation.
Furthermore, whenever we encounter a product $q_i d_i$ it will be reduced to 1.
By choosing a suitable monomial ordering, one can control further features of the reduced form as will be discussed in more detail in Section \ref{sec:blockorder} and Appendix\ref{sec:leinartaspolyred}.
Here, we only note that for any monomial ordering, which sorts first for the $q_i$ and then for the $x_i$, denominators with disjoint zeros will be separated and it is justified to call the reduction a partial fraction decomposition.
Using such a monomial ordering, we thus achieve a multivariate partial fraction decomposition with the following two steps: 1. calculate the Gr\"obner basis of the ideal $I$ and 2. reduce the rational function with respect to this Gr\"obner basis.
This reduction yields a unique remainder, which is the partial fractioned form of $r$.
A complete algorithm to bring a rational function into a unique partial fractioned form can be formulated as follows (see also \cite{Abreu:2019odu}).

\begin{alg}
\label{alg:mapart}
\emph{Multivariate partial fraction decomposition.}
A rational function $r(x_1,\ldots)\in K[x_1,\ldots]$ can be decomposed into partial fractions using the following steps.
\begin{enumerate}
% step 1
    \item Bring the rational function into the form $n(x_1,\ldots)/d(x_1,\ldots)$ and cancel common factors in $n$ and $d$ such that they are coprime.
% step 2
    \item Factorize $d$ over $K$.
    Let's call the irreducible factors of the denominator $d_i(x_1,\ldots)$ for $i=1,\ldots,m$.
% step 3
    \item For each denominator factor $d_i(x_1,\ldots)$ introduce a new indeterminant $q_i$ which represents the inverse of $d_i(x_1,\ldots)$, i.e.\ $q_i = 1/d_i(x_1,\ldots)$.
    Express all denominators in the problem in terms of the $q_i$ such that the rational function becomes a polynomial $p \in K[q_1,\ldots,x_1,\ldots]$.
% step 4
    \item Calculate the Gr\"obner basis of the ideal $I$ generated by
     $\{q_1d_1(x_i)-1,$ $\ldots,$ $q_m d_m(x_i)-1\}$ using a monomial ordering which sorts first for the $q_i$ and then for the $x_i$.
% step 5
    \item Find the fully reduced form of the polynomial $p$ with respect to the Gr\"obner basis.
% step 6
    \item Replace back $q_i \rightarrow 1/d_i$.
\end{enumerate}
\end{alg}

Step 1 eliminates spurious poles and ensures a unique final form.
When operating on large sums, it may be useful to skip this step and operate on the invidual terms of the sum separately.
In this situation, one can still arrive at a unique output for the sum when considering the decomposition of each term separately, but using the global set of denominators in the sum.
For instance, for the example of Section \ref{sec:leinartas}, Algorithm\ref{alg:mapart} finds a unique form independent of the specific initial form of $r$ (Eqs. \eqref{eq:red1} and \eqref{eq:red2}) once all possible denominators in the problem are specified.
The difference with respect to the standard {\Leinartas} decomposition lies in the fact, that we specify by the monomial ordering which denominators are preferred in the partial fractioned result.
The elimination of denominators which are known to be spurious can then be achieved by a suitable choice of the monomial ordering; this will be discussed in the next Section.

Step 6 is optional, of course. Leaving the denominators in abbreviated form can allow for more compact representations and as a preparation for an efficient numerical implementation.

%%%%%%%%%%%%%%%%%%%%%%%%%%%%%%%%%%%%%%%%%%%%%%%%%%%%%%%%%%%%%%%%%%%%%%%%%%%%%%%%%%%%%%%%
\subsection{Monomial ordering}
\label{sec:blockorder}
%%%%%%%%%%%%%%%%%%%%%%%%%%%%%%%%%%%%%%%%%%%%%%%%%%%%%%%%%%%%%%%%%%%%%%%%%%%%%%%%%%%%%%%%

The choice of monomial ordering has a crucial impact on the properties of the output of Algorithm~\ref{alg:mapart} and the performance of its computation.
We show in Appendix\ref{sec:leinartaspolyred} that the output form will satisfy {\Leinartas}' requirement (i) and thus separate denominators with disjoint zeros as long as the monomial ordering orders all $q_i$ before the $x_i$, while {\Leinartas}' requirement (ii) may or may not be satisfied depending on further details.

The calculation of the Gr\"obner basis can be very challenging for practical applications of Algorithm~\ref{alg:mapart}, depending heavily on the choice of monomial ordering.
Here we propose a specific monomial ordering which aims to provide good computational performance.
We tested this ordering in various calculations of scattering amplitudes, and we were able to compute the Gr\"obner basis and partial fraction the required rational functions in all cases that we considered.

To motivate our choice of monomial ordering, let us go back to Eq.\ \eqref{Eq:iterativePF}. Note that the prefactors in front of the denominators $x-f(y)$ and $x-g(y)$ do not depend on $x$. In the case where now $f(y)-g(y)$ is itself a valid denominator, this identity would therefore actually be a valid replacement rule and corresponds to a polynomial in the ideal $I$ in Eq.\ \eqref{Eq:ideal}. To ensure that this polynomial leads to the reduction \eqref{Eq:iterativePF}, one can choose a monomial ordering, in which the inverse denominator variables $q_1$ and $q_2$ corresponding to $d_1=x-f(y)$ and $d_2=x-g(y)$, respectively, are ``greater'' than the inverse denominator variable $q_3$ corresponding to $d_3=f(y)-g(y)$. This motivates the use of a block ordering, in which we group all denominators depending on $x$ and $y$ and all denominators depending on only $x$ or only $y$.

\begin{alg}
\label{alg:ordering}
\emph{Monomial block ordering.} A suitable monomial ordering for the ring $K[q_1,\ldots,q_m,x_1,\ldots,x_n]$ in Algorithm~\ref{alg:mapart} can be constructed as follows.
\begin{itemize}
    \item[1.] Group the denominators $d_1,\ldots,d_m$ by their dependence on all variables $x_1,\ldots,x_n$, such that denominators which depend on the same set of variables form a group; each group will correspond to a block in the monomial ordering.
    \item[2.] Sort the groups according to the number of variables they depend on; a group with denominators depending on fewer variables is considered ``smaller'' than a group with denominators depending on more variables.
    \item[3.] In each group, sort the denominators according to their total degree.
    \item[4.] Replace the denominators $d_i$ by the corresponding inverse denominator variables $q_i$ and add a last group containing the variables $x_1,\ldots,x_n$.
    \item[5.] Let the sequence of groups of variables define the blocks of a monomial ordering. Within each block, use the ``standard'' degree reverse lexicographic ordering (degrevlex).
\end{itemize}
\end{alg}

\noindent
Let us consider at an example with the denominators
\begin{equation}
    d_1 = x^2+y,\; d_2 = x-y,\; d_3 = x+1,\; d_4 = x^2-3,\; d_5 = y+1,\; d_6 = y\,.
\end{equation}
The ideal $I$ is generated by the polynomials
\begin{equation}\label{eq:ideal_example}
    \{q_1(x^2+y)-1,q_2(x-y)-1,q_3(x+1)-1,q_4(x^2-3)-1,q_5(y+1)-1,q_6y-1\}.
\end{equation}
In steps 1-3, we identify the three groups of denominators
\begin{equation}
        \{\{d_1 = x^2+y,\; d_2=x-y\},\{d_4=x^2-3,\; d_3=x+1\},\{d_5=y+1,\; d_6=y\}\},
\end{equation}
which gives in step 4
\begin{equation}
    \{\{q_1,q_2\},\{q_4,q_3\},\{q_5,q_6\},\{x,y\}\}.
\end{equation}
This defines a monomial ordering with 4 blocks, see step 5, which would be used to calculate the Gr\"obner basis and the subsequent polynomial reductions.

Our proposal for the monomial ordering aims to reduce the computational effort of the decomposition and to prefer low degrees of the denominator and numerator polynomials.
In our experiments, it allowed for a significantly faster calculation of the Gr\"obner basis than a global degree ordering or a lexicographical ordering.
We would like to point out that our choice of monomial order will in general only guarantee {\Leinartas}' requirement (i) but not (ii).
A lexicographical ordering could achieve a {\Leinartas} decomposition fulfilling both requirements, but potentially at the price of significantly increased polynomial degrees.
We give details and an example in Appendix\ref{sec:leinartaspolyred}.

%%%%%%%%%%%%%%%%%%%%%%%%%%%%%%%%%%%%%%%%%%%%%%%%%%%%%%%%%%%%%%%%%%%%%%%%%%%%%%%%%%%%%%%%
\subsection{Example for the algorithm}
\label{sec:example}
%%%%%%%%%%%%%%%%%%%%%%%%%%%%%%%%%%%%%%%%%%%%%%%%%%%%%%%%%%%%%%%%%%%%%%%%%%%%%%%%%%%%%%%%

We consider again the rational expression in Eq.\ \eqref{eq:rexampleambig},
\begin{equation}
    r=\frac{2y-x}{y(x+y)(y-x)},
    \label{eq:rexample1}
\end{equation}
and apply our partial fraction algorithm to it. We identify $3$ irreducible denominators, and therefore consider the ideal
\begin{equation}
I = \left< q_1(x-y)-1,q_2y-1,q_3(x+y)-1\right> .
\end{equation}
Our monomial ordering is defined by
\begin{equation}
    \{\{q_3, q_1\},\;\; \{q_2\},\;\; \{x, y\}\}.
\end{equation}
and the resulting Gr\"obner basis is obtained as
\begin{equation}
    g=\{-1+q_2y,-1+q_1x-q_1y,-1+q_3x+q_3y,-q_1 q_2+2q_1 q_3+q_2 q_3\}.
\end{equation}
Reducing $r=(2x-y)q_1q_2q_3$ with respect to $g$ yields the reduced form
\begin{equation}\label{eq:red_g}
    r=-\frac{1}{2}q_1 q_2+\frac{3}{2} q_2 q_3=\frac{3}{2y(x+y)}+\frac{1}{2y(y-x)},
\end{equation}
which is our partial fractioned result.
Lets now assume we start with the equivalent expression from Eq.\ \eqref{eq:red2}, i.e.\
\begin{equation}
r=\frac{1}{y(x + y)} + \frac{1}{(x - x)(x + y)}=q_1 q_2+q_2 q_3.
\end{equation}
Reducing this expression with respect to $g$ indeed yields exactly the same result as before, given by Eq.\ \eqref{eq:red_g}. Therefore, in this example, our algorithm recognizes that both input forms of $r$ are equivalent, and gives a unique result.
If we however start in another form, in which spurious denominators occur, one has to be careful.

Let us consider a different representation of the same rational expression $r$ in \eqref{eq:rexample1}, this time containing an additional spurious denominator:
\begin{equation}
    r=\frac{1}{y(x + y)} +  \frac{1}{2x}\frac{1}{(y-x)}-\frac{1}{2x}\frac{1}{(x+y)}.
\end{equation}
In this case we would identify four denominators and consider the ideal
\begin{equation}
I = \left< q_1(x-y)-1,q_2y-1,q_3(x+y)-1,q_4x-1\right>.
\end{equation}
Our method produces in this case the block ordering
\begin{equation}
    \{\{q_3, q_1\},\;\; \{q_2\},\;\;\{q_4\},\;\; \{x, y\}\}.
\end{equation}
Calculating the corresponding Gr\"obner basis and reducing with respect to that basis yields
\begin{equation}
   r = -\frac{3}{2}q_3 q_4 -\frac{1}{2}q_1 q_4 + q_2 q_4\,.
\end{equation}
We see that the spurious inverse denominator $q_4$ does not drop out, since we use a monomial ordering in which $q_4 \prec q_1,q_2,q_3$. However, if we would have chosen a different monomial ordering, in which $q_4$ is greater than all other $q_i$, the spurious denominator would drop out after reduction.
In this way, one can locally eliminate denominators which are known to be spurious or add additional denominators to the problem without altering previous reductions.
These features are also supported by our package, cf.\ Section \ref{sec:package}.

%%%%%%%%%%%%%%%%%%%%%%%%%%%%%%%%%%%%%%%%%%%%%%%%%%%%%%%%%%%%%%%%%%%%%%%%%%%%%%%%%%%%%%%%
\subsection{Efficient reduction of factorized inputs}
\label{sec:tuning}
%%%%%%%%%%%%%%%%%%%%%%%%%%%%%%%%%%%%%%%%%%%%%%%%%%%%%%%%%%%%%%%%%%%%%%%%%%%%%%%%%%%%%%%%

If the input expression is in a form with a single common denominator, i.e.\ 
\begin{equation}
    r=q_1^{\alpha_1}...q_m^{\alpha_m}\times \mathcal{N},
\label{eq:rfactorized}
\end{equation}
where $ \mathcal{N}$ is a possibly lengthy numerator, it may be optimal from a performance perspective to just fully expand the polynomial prior to reduction.
In this case, we propose to the following guided scheme.

\begin{alg}
\emph{Iterated reductions.} The polynomial reduction of $r$ in Eq.\ \eqref{eq:rfactorized} with respect to $I$ can be performed as follows.
\begin{enumerate}
    \item Set $p = \mathcal{N}$ and $Q=\{q_1^{\alpha_1},\ldots,q_m^{\alpha_m}\}$.
    \item Identify the ``smallest'' inverse denominator factor $q_i$ in $Q$ according to the monomial ordering defined in Section \ref{sec:blockorder}. Set $p \leftarrow  p\times q_i^{\alpha_i}$ and remove $q_i^{\alpha_i}$ from $Q$.
    \item Replace $p$ by its reduced form w.r.t.\ the Gr\"obner basis of the ideal $I$.
    \item If $Q$ is non-empty, goto step 2. Otherwise, stop and return $p$.
\end{enumerate}
\end{alg}

The reason why this reduction scheme can be much faster than a ``naive'' direct reduction is, that the decomposed form of the common denominator $q_1^{\alpha_1},\ldots,q_m^{\alpha_m}$ alone can result in huge expression, which only shortens once the numerator is taken into account. The iterative reduction of one denominator at a time avoids this intermediate expression swell, because it avoids the clustering of many denominators at one reduction step.

In the {\mathematica} functions of our package, the iterated scheme above can be selected as an option. Our {\form} implementation of the polynomial reductions uses this scheme by default.
We observed an additional speed-up in our implementation in {\mathematica} by partitioning the whole expression into smaller pieces and reducing these individually in step 2. An example with a comparison of timings for different sizes of these partitions can be found in Section \ref{sec:example_5_point}.

%%%%%%%%%%%%%%%%%%%%%%%%%%%%%%%%%%%%%%%%%%%%%%%%%%%%%%%%%%%%%%%%%%%%%%%%%%%%%%%%%%%%%%%%
\subsection{A comment on rational reconstructions}
\label{sec:ratrecon}
%%%%%%%%%%%%%%%%%%%%%%%%%%%%%%%%%%%%%%%%%%%%%%%%%%%%%%%%%%%%%%%%%%%%%%%%%%%%%%%%%%%%%%%%

Symbolic manipulations of large rational expressions can be computationally expensive.
Finite field sampling and rational reconstruction techniques allow to prevent intermediate expression swell and have become quite standard in high energy physics calculations today \cite{vonManteuffel:2014ixa,vonManteuffel:2016xki,Peraro:2016wsq,Peraro:2019svx,Klappert:2019emp,Smirnov:2019qkx}.
The basic idea is to set the indeterminates to numerical values in prime fields, perform all complicated manipulations with machine-sized integers instead of rational functions, and to reconstruct the final rational function of interest from many such samples.
In this Section we would like to discuss the usage of information about the denominator structure for an improved reconstruction of rational functions.
A major goal is to reduce the number of samples needed to reconstruct the rational functions, thus speeding up the computation.
Furthermore, in applications involving linear relations between Feynman integrals, the partial fractioned forms are often particularly simple and it would be a significant advantage to directly reconstruct the partial fractioned form.

We first start with the following observation concerning the prediction of denominator factors in rational functions, which are to be reconstructed from numerical samples.
Linear relations (integration-by-parts identities) between Feynman integrals typically involve only a small set of denominator factors.
Moreover, choosing appropriate basis or master integrals can help to avoid spurious denominators in the calculation.
In practice, the analysis of the denominator factors may be performed by considering only a small set of linear relations for a specific sector (set of distinct propagators) and setting subsector integrals (with fewer distinct propagators) to zero \cite{Smirnov:2020quc,Usovitsch:2020jrk}.
This results in a list of possible denominator factors to be expected for a specific rational function.
This knowledge helps to predict the denominator structure of the multivariate rational function e.g.\ by univariate reconstructions.

Here, we propose a new method, which allows to straight-forwardly guess denominator factors and their powers for a rational function, based on a list of candidate denominator factors and a small number of numerical probes (samples) of the rational function to reconstruct.
Once the denominator is known or at least partial information about it, the full rational function can be reconstructed from (substantially) fewer samples.
Let us assume we analyzed our rational functions and expect a list of irreducible denominator factors $d_1,\ldots,d_m$ in the result.
We assume that the coefficient to be reconstructed is of the form 
\begin{equation}
r = \frac{\mathcal{N}}{d_1^{\alpha_1} \cdots d_m^{\alpha_m}},
\label{eq:ansatz}
\end{equation}
where $\mathcal{N}$, $d_1$, \ldots, $d_m$ are polynomials in $\mathbbm{Q}[x_1,\ldots,x_n]$, and $\alpha_1,\ldots \in \mathbbm{N}_0$.

\begin{recipe}\emph{Denominator guess.} For a rational function, which is expected to be of the form \eqref{eq:ansatz} with known denominator factors $d_1(x_1,\ldots),\ldots$, the following construction provides a guess for the full denominator. 
\begin{enumerate}
    \item Find integer samples for all $x_1,\ldots,x_n$ such that all $d_1$, \ldots, $d_m$ have a distinct, largest prime factor, which we denote by $p_i$, and each of these largest prime factors has multiplicity one. This requirement can be weakened but the stated form simplifies the following analysis.
    \item Evaluate the rational function for the numerical values of the $x_1,\ldots,x_n$. In practice, this may mean e.g.\ linear system solving for several prime fields, Chinese remaindering and rational reconstruction.
    The result of this step is a number $r_{\text{num}}\in\mathbbm{Q}$, which can be written in the form $r_{\text{num}}=n_{\text{num}}/d_{\text{num}}$ with $n_{\text{num}},d_{\text{num}} \in \mathbbm{Z}$.
    \item Perform a prime factor decomposition of the integer number $d_{\text{num}}$. Due to Eq.\ \eqref{eq:ansatz}, each factor $d_i^{\alpha_i}$ must contribute $p_i^{\alpha_i}$ to the prime factors in $d_{\text{num}}$. On the other hand, if we find a factor $p_i^{\alpha_i}$, we can take it as an indication that it was generated by $d_i^{\alpha_i}$ in the denominator. In this way, we can make a guess for the denominator just by counting the multiplicities of the $p_i$ in the factorization of $d_{\text{num}}$.
\end{enumerate}
\end{recipe}

In step 3 our guess may be off due the presence of a new denominator factor which was not in the candidate list or a factor $p_i$ appearing in the coefficients of the numerator polynomial $\mathcal{N}$.
One could imagine to repeat the above construction for several samples of the indeterminates and combine this information to validate or correct the guess.
In our experiments we have just picked somewhat larger integers and were in fact not even faced with these problems for the cases that we checked.

The above method lets us therefore identify (parts of) the denominator of the expression we want to reconstruct.
A robust way to use this guess for the full reconstruction of a rational function is to multiply all samples with this denominator and reconstruct the resulting expression as if it was a rational function.
If the guess was accurate, only the numerator needs to be reconstructed, thereby reducing the number of required samples.
We emphasize that this method will allow for a successful reconstruction of the result even if the guess was incomplete or inaccurate.
After reconstruction one can perform the partial fractioning as discussed in Section \ref{sec:alg}.

We successfully applied this method to integration-by-parts reductions of Feynman integrals for problems with up to three loops.
Considering the reconstruction of just a single variable, we typically observe a reduction of the number of samples by a factor of about 2 due to the guessed denominators.
For a reduction problem with five variables, the iterated reconstruction of all variables resulted in an overall reduction of the number of samples by about a factor of 10.

As a next step, one can ask if it is possible to directly reconstruct the result in partial fractioned form. Indeed, assuming one knows all irreducible denominator factors of an expression beforehand, our method in Section~\ref{sec:alg} actually allows to systematically construct a basis of all monomials that can be expected in the final partial fractioned result, using a bound on the monomial degree in the construction.
This, in principle, allows to match against an ansatz and reconstruct the partial fractioned form directly.
We observed in our experiments that depending on the exact strategy the ansatz in partial fractioned form may require more samples to fix the free coefficients than the reconstruction in the common denominator representation along the lines described above.
Furthermore, an incomplete list of denominator factors would prevent a successful fit of a partial fractioned ansatz as well as an insufficient degree of the ansatz.
In contrast, the method for the reconstruction in common-denominator form presented above is robust with respect to incomplete denominator information.

%%%%%%%%%%%%%%%%%%%%%%%%%%%%%%%%%%%%%%%%%%%%%%%%%%%%%%%%%%%%%%%%%%%%%%%%%%%%%%%%%%%%%%%%
\section{The package}
\label{sec:package}
%%%%%%%%%%%%%%%%%%%%%%%%%%%%%%%%%%%%%%%%%%%%%%%%%%%%%%%%%%%%%%%%%%%%%%%%%%%%%%%%%%%%%%%%
\subsection{Installation and basic usage}
\label{sec:usage}
%%%%%%%%%%%%%%%%%%%%%%%%%%%%%%%%%%%%%%%%%%%%%%%%%%%%%%%%%%%%%%%%%%%%%%%%%%%%%%%%%%%%%%%%

We implement the described algorithm in a {\mathematica} package, that can be downloaded as follows:
\begin{equation*}
\texttt{git clone https://gitlab.msu.edu/vmante/multivariateapart.git}
\end{equation*}
The package consists of just a single file, \texttt{MultivariteApart.wl}, which could also be downloaded individually.
We recommend to place it in a standard Mathematica packages directory, e.g.\ \texttt{\textapprox/.Mathematica/Applications}, such that it can be loaded with
\small
\begin{mmaCell}{Code}
Needs["MultivariateApart`"]
\end{mmaCell}
\normalsize
from any directory. Alternatively, it can always be loaded by specifying the full path to the file.
The main function, which implements the algorithm described in Section \ref{sec:alg} can be called by  \textmma{MultivariateApart}. Using this command on the example of Section \ref{sec:example} yields:
\small
\begin{mmaCell}{Code}
\mmaDef{MultivariateApart}[(2y-x)/(y(x+y)(y-x))]
\end{mmaCell}
\begin{mmaCell}{Output}
- \mmaFrac{1}{2(x-y)y} + \mmaFrac{3}{2y(x+y)}
\end{mmaCell}
\normalsize
The function provides a unique representation and may be all that is needed in many cases.

It can be convenient to keep the inverse factors abbreviated for further processing, which can be achieved by using \textmma{MultivariateAbbreviatedApart} instead:
\small
\begin{mmaCell}{Code}
\mmaDef{MultivariateAbbreviatedApart}[(2y-x)/(y(x+y)(y-x))] // Normal
\end{mmaCell}
\begin{mmaCell}{Output}
\{-((q1 q2)/2) + (3 q2 q3)/2, \{q1 -> 1/(x - y), q2 -> 1/y, 
  q3 -> 1/(x + y)\}\}
\end{mmaCell}
\normalsize
Here, \textmma{Normal} was used to convert the dispatch table in the second element of the returned list into normal form.
One can pass a list of previously identified denominator factors as a second argument to \textmma{MultivariateAbbreviatedApart}:
\small
\begin{mmaCell}{Code}
\mmaDef{MultivariateAbbreviatedApart}[(2y-x)/(y(x+y)(y-x)), {x-y,y,x+y}]
\end{mmaCell}
\normalsize
This allows one to work with a global list of abbreviations across several expressions.

The package provides further functions and options, which may be useful for more granular control, performance tuning and interfacing with other computer algebra systems.
We will describe some of this functionality in the following.
Please see the output of \textmma{? MultivariateApart`*} for a full list of available functions and options.

%%%%%%%%%%%%%%%%%%%%%%%%%%%%%%%%%%%%%%%%%%%%%%%%%%%%%%%%%%%%%%%%%%%%%%%%%%%%%%%%%%%%%%%%
\subsection{Advanced usage pattern}
%%%%%%%%%%%%%%%%%%%%%%%%%%%%%%%%%%%%%%%%%%%%%%%%%%%%%%%%%%%%%%%%%%%%%%%%%%%%%%%%%%%%%%%%

The ``all-in-one'' functions described so far internally call a sequence of other functions, which can also be used directly for more advanced use cases.
With default options, the first step is a call to
\textmma{Together} to eliminate spurious denominators.
This behavior as well as other details can be configured through the options of the function. Of course, for our example this step is irrelevant.
Next, \textmma{AbbreviateDenominators} is called, which finds all denominators in an expression and replaces them by a predefined symbol (the default is $q_i$).
\small
\begin{mmaCell}{Code}
{expr, dens, qis} = AbbreviateDenominators[(2y-x)/(y(x+y)(y-x)]
\end{mmaCell}
\begin{mmaCell}{Output}
\{-q1 q2 q3 (-x + 2y), \{x - y, y, x + y\}, \{q1, q2, q3\}\}
\end{mmaCell}
\normalsize
The output is a list, whose first entry is the expression in terms of the denominator symbols, the second entry a list of all irreducible denominator factors and the last entry a list of their abbreviations.
Optionally, \textmma{AbbreviateDenominators} accepts a list of denominator factors as a second argument to support working with a global list of abbreviations.
Next, \textmma{ApartOrder} is used to define the monomial order:
\small
\begin{mmaCell}{Code}
ord = ApartOrder[dens, qis]
\end{mmaCell}
\begin{mmaCell}{Output}
\{\{q3, q1\}, \{q2\}, \{x, y\}\}
\end{mmaCell}
\normalsize
The returned list represents the monomial order, that will be used for the Gr\"obner basis computation, cf.\ Section \ref{sec:blockorder}.
The Gr\"obner basis is obtained with
\small
\begin{mmaCell}{Code}
gb = ApartBasis[dens, qis, ord]
\end{mmaCell}
\begin{mmaCell}{Output}
\{-1 + q2 y, -1 + q1 x - q1 y, -1 + q3 x + q3 y,
 -q1 q2 + 2 q1 q3 + q2 q3\}
\end{mmaCell}
\normalsize
The actual reduction or partial fractioning is performed with
\small
\begin{mmaCell}{Code}
ApartReduce[expr, gb, ord]
\end{mmaCell}
\begin{mmaCell}{Output}
- \mmaFrac{q1 q2}{2} + \mmaFrac{3 q2 q3}{2}
\end{mmaCell}
\normalsize

This example also shows the alternative way of using our package. Instead of using \texttt{MultivariateApart} on a single expression, we can think about the situation where we know all denominators of a set of expressions beforehand, and then want to bring a give expression into a unique form. Lets assume we have an additional denominator $x$ in the problem. We can then first calculate the Gr\"obner base including this additional denominator and then reduce with respect to that base. As we have seen already in Section \ref{sec:example} there is no guarantee that an expression is free of spurious singularities in this case:
\small
\begin{mmaCell}{Code}
{dens, qis} = {{x - y, y, x + y, x}, {q1, q2, q3, q4}};
ord = ApartOrder[dens, qis]
\end{mmaCell}
\begin{mmaCell}{Output}
\{\{q3, q1\}, \{q2\}, \{q4\}, \{x, y\}\}
\end{mmaCell}
\normalsize
Note that the algorithm considers $q_4=1/x$ to be ``simple'' and will try to eliminate the other $q_i$ in favor of $q_4$.
\small
\begin{mmaCell}{Code}
gb = ApartBasis[dens, qis, ord];
ApartReduce[expr, gb, ord]
\end{mmaCell}
\begin{mmaCell}{Output}
- \mmaFrac{q1 q4}{2} + q2 q4 - \mmaFrac{3 q3 q4}{2}
\end{mmaCell}
\normalsize
Here, the pole $q_4=1/x$ is actually spurious.

If we knew already beforehand that $q_4$ is spurious, we could have used a monomial order such that $q_4$ drops out (one could think of a situation, in which one knows that some denominators should not occur in the final answer). In this example, this can be done by changing the list \textmma{ord} in the following way: we just reorder the blocks in \textmma{ord} by moving the block with $q_4$ to the left most position (i.e.\ that block is greater than all other): 
\small
\begin{mmaCell}{Code}
ord = {{q4}, {q3, q1}, {q2}, {x, y}}
gb = ApartBasis[dens, qis, ord];
ApartReduce[expr, gb, ord]
\end{mmaCell}
\begin{mmaCell}{Output}
- \mmaFrac{q1 q2}{2} + \mmaFrac{3 q2 q3}{2}
\end{mmaCell}
\normalsize
As expected, the result is free of the spurious pole $q_4$ and we find the same answer, as using \textmma{MultivariateApart} directly on the whole expression.
Let us demonstrate the important fact that, once a monomial order is chosen, the final answer is essentially insensitive to the exact input form, and one can equally well operate on individual terms of a sum:
\small
\begin{mmaCell}{Code}
terms = ApartReduce[-q1 q4/2, gb, ord] +
 ApartReduce[q2 q4, gb, ord] + 
 ApartReduce[-3 q3 q4/2, gb, ord]
\end{mmaCell}
\begin{mmaCell}{Output}
q2 q4 + 3/2 (q2 q3 - q2 q4) + 1/2 (-q1 q2 + q2 q4)
\end{mmaCell}
\normalsize
Note how in this case the partial fractioned form of the invidual terms actually exhibit the pole $q_4$, but they cancel in the sum after a call to \textmma{Expand},
\small
\begin{mmaCell}{Code}
Expand[terms]
\end{mmaCell}
\begin{mmaCell}{Output}
- \mmaFrac{q1 q2}{2} + \mmaFrac{3 q2 q3}{2}
\end{mmaCell}
\normalsize
which is the same result as before.

%%%%%%%%%%%%%%%%%%%%%%%%%%%%%%%%%%%%%%%%%%%%%%%%%%%%%%%%%%%%%%%%%%%%%%%%%%%%%%%%%%%%%%%%
\subsection{Interface with Singular}
%%%%%%%%%%%%%%%%%%%%%%%%%%%%%%%%%%%%%%%%%%%%%%%%%%%%%%%%%%%%%%%%%%%%%%%%%%%%%%%%%%%%%%%%

Although our package can be used entirely in {\mathematica}, for difficult examples we recommend to perform the computation of the Gr\"obner basis in {\singular}. An input file for the calculation of the Gr\"obner basis can be generated with the function \texttt{WriteSingularBasisInput}, which can be executed in {\singular} with the \texttt{read()} function. As an example how to use {\singular} with our package we return to the previous example:

\small
\begin{mmaCell}{Code}
{expr, dens, qis} = AbbreviateDenominators[(2y-x)/(y(x+y)(y-x)]
\end{mmaCell}
\begin{mmaCell}{Output}
\{-q1 q2 q3 (-x + 2y), \{x - y, y, x + y\}, \{q1, q2, q3\}\}
\end{mmaCell}
\begin{mmaCell}{Code}
ord = ApartOrder[dens, qis]
\end{mmaCell}
\begin{mmaCell}{Output}
\{\{q3, q1\}, \{q2\}, \{x, y\}\}
\end{mmaCell}
\normalsize
We can now prepare the Gr\"obner basis calculation with the command
\small
\begin{mmaCell}{Code}
WriteSingularBasisInput[dens, qis, ord, "dir"]
\end{mmaCell}
\normalsize
Here, \texttt{"dir"} denotes the name of the directory to which input and output files are written.
The specified directory will be created if it does not exist.
All commands for {\singular} are stored in the input file \texttt{apartbasisin.sing}.
In {\singular} we can use \texttt{execute(read(apartbasisin.sing))} to execute these commands.
After execution, the file \texttt{apartbasisout.m} is created, which contains the Gr\"obner basis. We can load it in {\mathematica}:
\small
\begin{mmaCell}{Code}
Get["apartbasisout.m"]
\end{mmaCell}
\begin{mmaCell}{Output}
\{-1 + q2 y, -1 + q1 x - q1 y, -1 + q3 x + q3 y,
 -q1 q2 + 2 q1 q3 + q2 q3\}
\end{mmaCell}
\normalsize
This is the same result as before when we used \texttt{ApartBasis} in the previous Section.

We note that {\singular} provides support for polynomial reductions and can therefore also be used instead of our function \textmma{ApartReduce}.
We observed good timings in this approach for some of our more complicated examples.

%%%%%%%%%%%%%%%%%%%%%%%%%%%%%%%%%%%%%%%%%%%%%%%%%%%%%%%%%%%%%%%%%%%%%%%%%%%%%%%%%%%%%%%%
\subsection{Interface with Form}
%%%%%%%%%%%%%%%%%%%%%%%%%%%%%%%%%%%%%%%%%%%%%%%%%%%%%%%%%%%%%%%%%%%%%%%%%%%%%%%%%%%%%%%%

We emphasize that the polynomial reduction step of our algorithm allows a straight forward implementation in symbolic manipulation programs such as {\form}, in which one uses \texttt{identify} statements to ``locally'' replace parts of expressions, that is, one considers one term of a potentially large sum at a time.
Having calculated the Gr\"obner basis, these \texttt{identify} statements can be derived by considering the leading term of each generator and generating rules that replace these leading terms.
We have successfully used our methods in this way and added the function \texttt{WriteFormProcedure} in our package. It can be invoked by
\small
\begin{mmaCell}{Code}
WriteFormProcedure[gb, ord, "dir"]
\end{mmaCell}
\normalsize
Here, \texttt{gb} is the Gr\"obner base from which the replacement rules are derived, \texttt{ord} is the monomial order that was used and \texttt{"dir"} is the directory where the {\form} procedure shall be written to. The command generates a {\form} input file called
\texttt{"apartreduce.h"}, containing a procedure that implements the reduction scheme described in Section \ref{sec:tuning}, and two auxiliary files, which have to be in the same directory. For usage in {\form}, the file has to be included in the beginning of a {\form} script with \texttt{"\#include apartreduce.h"} and the reduction algorithm can be invoked by
\texttt{"\#call apartreduce(expr)"}, where \texttt{expr} is a local {\form} expression. Note that \texttt{"apartreduce.h"} contains the definition of the denominator symbols in a specific order, which is needed for an efficient reduction. The user should therefore \emph{not define these symbols} at another point in his or her {\form} scripts.

It is also possible to use {\form} as a back end for the computation of polynomial reductions in \textmma{MultivariateApart}, \textmma{MultivariateAbbreviatedApart}, and \textmma{ApartReduce} by specifying the option \texttt{UseFormProgram -> True}.
For example, the command 
\small
\begin{mmaCell}{Code}
ApartReduce[expr, gb, ord, UseFormProgram->True]
\end{mmaCell}
\normalsize
calls the external program {\form} to perform the actual computation.
Here, the {\form} executable is assumed to be in the user's path, and the directory \textmma{\$ApartTemporaryDirectory} will be used for all temporary files.

%%%%%%%%%%%%%%%%%%%%%%%%%%%%%%%%%%%%%%%%%%%%%%%%%%%%%%%%%%%%%%%%%%%%%%%%%%%%%%%%%%%%%%%%
\section{Applications}
\label{sec:applications}
%%%%%%%%%%%%%%%%%%%%%%%%%%%%%%%%%%%%%%%%%%%%%%%%%%%%%%%%%%%%%%%%%%%%%%%%%%%%%%%%%%%%%%%%

The methods described in this paper have been successfully applied to several calculations of Feynman amplitudes. In this Section, we provide some details for these applications.

%%%%%%%%%%%%%%%%%%%%%%%%%%%%%%%%%%%%%%%%%%%%%%%%%%%%%%%%%%%%%%%%%%%%%%%%%%%%%%%%%%%%%%%%
\subsection{One-loop amplitudes for \texorpdfstring{$\gamma^* \gamma^* \to l^+l^-$}{AA -> ll} with finite lepton mass}
\label{sec:AAll}
%%%%%%%%%%%%%%%%%%%%%%%%%%%%%%%%%%%%%%%%%%%%%%%%%%%%%%%%%%%%%%%%%%%%%%%%%%%%%%%%%%%%%%%%

One of us calculated all next-to-leading-order helicity amplitudes for the process
\begin{equation}
    \gamma^*(p_1)+ \gamma^*(p_2) \to l^+(p_3)+ l^-(p_4),
\end{equation}
where the quantities in parenthesis denote the four-momenta of the particles.
The calculation \cite{Heller:2019dyv,Heller:2020lnm} included the full dependence on the lepton mass using the ideas of this paper.
Defining the kinematic quantities
\begin{gather}
p_1^2=t_1,\quad p_2^2=t_2,\quad p_3^2=p_4^2=m^2,
\quad
(p_1+p_2)^2=s,\quad (p_1-p_3)^2=t,
\end{gather}
and setting $m^2\equiv 1$, one encounters a total of $15$ denominators, for which our package finds the optimized ordered form
\small
\begin{align}
    \{&\{s - 2\;s\;t + s^2\;t + s\;t^2 - s\;t_1 - s\;t\;t_1 + t_1^2 - s\;t_2 - s\;t\;t_2 - 2\;t_1\;t_2 + s\;t_1\;t_2 + t_2^2,\nonumber\\
 & 1 - 2\;s + s^2 - 2\;t + 2\;s\;t + t^2 - 4\;t_1 + 2\;t_2 - 2\;s\;t_2 - 2\;t\;t_2 + t_2^2,\nonumber\\
 &1 - 2\;s + s^2 - 2\;t + 2\;s\;t + t^2 + 2\;t_1 - 2\;s\;t_1 - 2\;t\;t_1 + t_1^2 - 4\;t_2, \nonumber\\
 &2 - s - t + t_1 + t_2, 1 - s - t + t_1 + t_2\}, \{s^2 - 2\;s\;t_1 + t_1^2 - 2\;s\;t_2 - 2\;t_1\;t_2 + t_2^2, \nonumber\\
 & s - t_1 - t_2\},
 \{1 - 2\;t + t^2 - 2\;t_2 - 2\;t\;t_2 + t_2^2\}, \{1 - 2\;t + t^2 - 2\;t_1 - 2\;t\;t_1 + t_1^2\}, \nonumber\\
 & \{t, -1 + t\}, \{s, 4 - s\}, \{t_2\}, \{t_1\}\}.
\end{align}
\normalsize
Using the block ordering described in Section \ref{sec:blockorder}, we can calculate the Gr\"obner basis for the ideal formed from these denominators using \texttt{slimgb} in {\singular} in a few minutes.

It is worth noting, that three of these denominators are spurious and result as an artefact of the basis of Lorentz invariant building blocks, that are used to build the amplitudes \cite{Tarrach:1975tu}. Namely, the three denominators that should drop out in the end are
\begin{equation}
t_1,\; t_2,\; s-t_1-t_2.
\end{equation}
Changing the monomial order slightly, such that these denominators form an extra block which is "greater" than all other denominators, we are still able to calculate the Gr\"obner basis of this ideal. We furthermore find that in the final answer these denominators indeed cancel out.

Comparing the size of the results in factorized form and partial fractioned form, we find a reduction by factor $7$ on average. However, some individual amplitudes are reduced by a factor of $20$ to $30$.

%%%%%%%%%%%%%%%%%%%%%%%%%%%%%%%%%%%%%%%%%%%%%%%%%%%%%%%%%%%%%%%%%%%%%%%%%%%%%%%%%%%%%%%%
\subsection{Two-loop amplitudes for \texorpdfstring{$gg \to ZZ$}{gg -> ZZ} with a closed top-quark loop}
%%%%%%%%%%%%%%%%%%%%%%%%%%%%%%%%%%%%%%%%%%%%%%%%%%%%%%%%%%%%%%%%%%%%%%%%%%%%%%%%%%%%%%%%

The algorithm presented in this work has also been applied to the calculation of the two-loop QCD corrections to the process
\begin{equation}
    g(p_1)+g(p_2) \to Z(p_3)+Z(p_4),
\end{equation}
involving a closed top-quark loop in Ref.~\cite{Agarwal:2020dye}.
In that calculation,
\begin{gather}
    p_1^2=p_2^2=0,\quad
    p_3^2=p_4^2=m_z^2,\quad
    (p_1+p_2)^2 = s,\quad
    (p_1-p_3)^2 = t,
\end{gather}
was used for the kinematics and the masses of the top quark and the $Z$ boson, $m_t$ and $m_z$, have been fixed by $m_t=1$ and $(m_z/m_t)^2=5/18$.
It is possible to choose a basis of master integrals, such that the dependence on the space-time dimension $d$ factorizes from denominators involving the kinematic invariants $s$ and $t$.
The coefficients contain the 9 $d$-dependent denominators
\begin{equation}
\{
5d+52,\;\; 3d-10,\;\; 3d-8,\;\; 2d-7,\;\; d-5,\;\; d-4,\;\; d-3,\;\; d-2,\;\; d  
\}
\end{equation}
and the 48 $d$-independent denominators
\begin{align}
&\{
\{
2125764 t^6+  8503056 s t^5
-5904900 t^5+ 12754584 s^2 t^4 -35901792 s t^4
\notag\\ &
+  2001105 t^4+ 8503056 s^3 t^3
-72275976 s^2 t^3+ 25850340 s t^3+  3863700 t^3
\notag\\ &
+ 2125764 s^4 t^2 -60466176 s^3 t^2
+  75149694 s^2 t^2 -19260180 s t^2 -797850 t^2
\notag\\ &
-18187092 s^4 t+  80752788 s^3 t
-73614420 s^2 t+ 17309700 s t -468000 t
\notag\\ &
+  29452329 s^4 -50490540 s^3
+ 25891650 s^2 -3676500 s+ 105625,\;\;
26244 t^6
\notag\\ &
+ 52488 s t^5 -14580 t^5+  26244 s^2 t^4
-224532 s t^4 -56295 t^4 -209952 s^2 t^3
\notag\\ &
+  116640 s t^3 + 32400 t^3+ 419904 s^2 t^2
-249480 s t^2+ 27900 t^2+  64800 s t
\notag\\ &
-18000 t + 2500,\;\;
524880 t^4 +  524880 s t^3 -583200 t^3+ 1889568 s^2 t^2
\notag\\ &
+ 1953720 s t^2+  243000 t^2
+ 3149280 s^2 t -1206900 s t -45000 t+ 145800 s^2
\notag\\ &
+  173250 s+ 3125,\;\;
104976 t^4+  314928 s t^3 -536544 t^3+ 209952 s^2 t^2
\notag\\ &
-1405512 s t^2+  398520 t^2
-839808 s^2 t+  599400 s t -106200 t -64800 s+ 9625,\;\;
\notag\\ &
104976 t^4
+ 104976 s t^3+  303264 t^3 -104976 s^2 t^2+  29160 s t^2 -301320 t^2
\notag\\ &
-104976 s^3 t
-244944 s^2 t+ 68040 s t+  88200 t+ 29160 s^3 -32400 s^2 -23400 s
\notag\\ &
-8375,\;\;
13122 s t^3+ 37908 t^3
+  26244 s^2 t^2+ 64881 s t^2 -56700 t^2+ 13122 s^3 t
\notag\\ &
+  26973 s^2 t -52650 s t+ 22725 t
+ 2025 s^2+  1800 s -2750,\;\;
104976 s^2 t^2
\notag\\ &
-898128 s t^2+  1454436 t^2+ 209952 s^3 t -1073088 s^2 t+ 64800 s t -808020 t
\notag\\ &
+  104976 s^4 -174960 s^3 -144180 s^2+ 51300 s+ 112225,\;\;
104976 s^2 t^2
\notag\\ &
-898128 s t^2+ 1454436 t^2 -839808 s^2 t+ 3841992 s t -808020 t
+  1679616 s^2
\notag\\ &
-997920 s+ 112225,\;\;
324 s^2 t^2 -2592 s t^2+ 5184 t^2+  648 s^3 t -2772 s^2 t
\notag\\ &
+ 1440 s t -2880 t
+  324 s^4 -180 s^3 -695 s^2 -200 s+ 400,\;\;
324 s^2 t^2 -2592 s t^2
\notag\\ &
+  5184 t^2 -2772 s^2 t+ 11808 s t -2880 t+  4489 s^2 
-3080 s + 400,\;\;
81 s t^2 -324 t^2
\notag\\ &
+ 162 s^2 t -414 s t+ 180 t+  81 s^3 -90 s^2+ 25 s -25,\;\;
81 s t^2 -324 t^2 -324 s t
\notag\\ &
+ 180 t -25,\;\;
3015 t^2+ 3015 s t -1675 t+ 2916 s^2 -6030 s+ 3350,\;\;
324 t^2
\notag\\ &
+ 648 s t -180 t+ 25,\;\;
324 t^2+ 558 s t -180 t+ 324 s^2 -155 s+ 25,\;\;
324 t^2
\notag\\ &
+ 324 s t -180 t -1296 s+ 25,\;\;
324 t^2+ 324 s t -180 t + 25,\;\;
324 t^2+  90 s t -180 t
\notag\\ &
+ 90 s^2 -25 s+ 25,\;\;
324 t^2 -180 t -324 s^2+  180 s+ 25,\;\;
162 t^2+  162 s t -135 t
\notag\\ &
+ 25,\;\;
18 t^2+ 18 s t -5 t + 5 s,\;\;
558 s t -90 t+ 558 s^2 -580 s +25,\;\;
558 s t -90 t
\notag\\ &
+ 180 s+ 25,\;\;
324 s t -90 t+  324 s^2 -270 s+ 25,\;\;
324 s t -90 t+ 25,\;\;
279 t
\notag\\ &
+ 279 s -200,\;\;
117 t+ 117 s -155,\;\;
18 t+ 36 s -5,\;\;
18 t+ 18 s -5,\;\;
18 t+ 9 s
\notag\\ &
-5,\;\;
18 t -18 s -5,\;\;
9 t+  9 s+ 31,\;\;
9 t + 9 s -5,\;\;
6 t + 6 s -5
\},
\{
324 s^2 -245 s
\notag\\ &
+ 90,\;\;
324 s -335,\;\;
18 s -5,\;\;
9 s + 134,\;\;
9 s -5,\;\;
9 s -10,\;\;
s - 4,\;\;
s
\},
\{
31 t + 5,%\;\;
\notag\\ &
18 t + 5,\;\;
18 t -5,\;\;
13 t + 10,\;\;
t -4,\;\;
t
\}\}.
\end{align}
During the course of the calculation, different choices for the master integrals were used, and the original choices did in fact introduce additional denominators, which mixed the dependence on $d$ and the kinematic invariants.
After changing to an improved set of master integrals, the methods presented in this paper allowed to systematically eliminate those denominators in a term-by-term approach and to very substantially reduce the size of the expressions.
In order to reduce the computational complexity, the partial fraction decomposition was performed for expressions expanded around $d=4$, allowing to effectively replace $d$ by 4 in the denominator analysis.
The actual polynomial reduction was performed directly in Singular in this case.

The final result for this amplitude was rather sizable.
However, the partial fractioned form of the rational coefficients of the master integrals allowed to conveniently generate fast code for their numerical evaluation.
Starting from a form with symbols for the inverse denominator factors, also powers of inverse denominator factors and of basic invariants were abbreviated.
In this way, each coefficient becomes a sum of simple products of a few factors each, where the basic factors (the powers) can be precomputed numerically.
This allows to straight-forwardly generate compiler friendly code, where each term is added to a summation variable numerically.

%%%%%%%%%%%%%%%%%%%%%%%%%%%%%%%%%%%%%%%%%%%%%%%%%%%%%%%%%%%%%%%%%%%%%%%%%%%%%%%%%%%%%%%%
\subsection{Two-loop amplitudes for massless five-particle scattering\label{sec:example_5_point}}
%%%%%%%%%%%%%%%%%%%%%%%%%%%%%%%%%%%%%%%%%%%%%%%%%%%%%%%%%%%%%%%%%%%%%%%%%%%%%%%%%%%%%%%%

For a comparison with the algorithm presented in Ref.\ \cite{Boehm:2020ijp}, we applied our algorithm to a benchmark example considered in that work: the integration-by-parts matrix provided in Ref.\ \cite{Bendle:2019csk} for the reduction of two-loop five-point functions in terms of a dlog basis.
The kinematic variables are chosen as 
\begin{equation*}
c_2 = s_{23}/s_{12},\quad c_3 = s_{34}/s_{12},\quad c_4 = s_{45}/s_{12},\quad c_5 = s_{15}/s_{12},
\end{equation*}
where $s_{12}=1$ and the $s_{ij}$ are Lorentz invariant functions of the external momenta.
Furthermore, $\epsilon$ is the parameter of dimensional regularization.
In the matrix of rational functions, we encounter a total of $24$ denominators\footnote{Note that the reduced matrix given in Ref.\ \cite{Boehm:2020ijp} has $32$ denominators because several appear twice modulo a sign change.},
\begin{align}
\{&-1 + c_3, c_3, c_2, 1 + c_2, c_2 + c_3, c_2 - c_4, 1 + c_2 - c_4, -1 + c_4, c_4, -1 + c_3 + c_4,\nonumber\\
&c_3 + c_4, c_2 - c_5, -1 + c_3 - c_5, c_2 + c_3 - c_5, c_2 - c_4 - c_5, 1 + c_2 - c_4 - c_5,\nonumber\\
&-1 + c_3 + c_4 - c_5, c_5, c_2^2 - 2c_2^2c_3 + c_2^2c_3^2 + 2c_2c_3c_4 - 2c_2c_3^2c_4 + c_3^2c_4^2 - 2c_2c_5 \nonumber\\
&+ 2c_2c_3c_5 + 2c_2c_4c_5 + 2c_3c_4c_5 + 2c_2c_3c_4c_5 - 2c_3c_4^2c_5 + c_5^2 - 2c_4c_5^2 + c_4^2c_5^2,\nonumber\\
&-1 + \epsilon, -1 + 2\epsilon, -1 + 3\epsilon,  -1 + 4\epsilon, 1 + 4\epsilon\}.
\end{align}

First, we apply \texttt{MultivariateApart} to the rational function appearing as the $16^{\text{th}}$ entry of the first row of the matrix and study the impact of different options on the runtime.
As can be seen in Table~\ref{tab:timing_coeff}, different options can alter the performance significantly despite producing the same output form.
In the Mathematica-only approach, enabling the iterated reduction scheme described in Section~\ref{sec:tuning} can help significantly to reduce the runtime, provided one selects a suitable finite partition size in addition.
We don't see a general argument why the partitioning should have the observed effect in all cases; we are tempted to speculate that finite internal buffer sizes in {\mathematica} could be the reason here.
Enabling the flag to use {\form} instead of {\mathematica} for the polynomial reductions results in the smallest runtime by quite a margin in this example.
We note that the {\form} implementation also uses iterated reduction approach of Section~\ref{sec:tuning}.

\begin{table}[t]
    \centering
    \begin{tabular}{|l|r|}
    \hline
    Option & Runtime  \\ 
    \hline
    \texttt{Iterate->False} \text{(default)} & 9457 sec \\
    \texttt{Iterate->True PartionSize->$\infty$} & 9254 sec\\
    \texttt{Iterate->True, PartionSize->5000} & 635 sec \\
    \texttt{Iterate->True, PartionSize->2000} & 422 sec \\
    \texttt{Iterate->True, PartionSize->1000} & 358 sec \\
    \texttt{Iterate->True, PartionSize->100} & 489 sec \\
    \texttt{UseFormProgram->True} & 60 sec \\
    \hline
    \end{tabular}
    \caption{Timing of \texttt{MultivariateApart} applied to entry $(1,16)$ of the matrix provided in \cite{Bendle:2019csk} for different options.
    \label{tab:timing_coeff}}
\end{table}

Next, we consider the partial fraction decomposition of all entries of the benchmark matrix in Ref.\ \cite{Bendle:2019csk}.
We ran our algorithm in two different setups, once by performing the partial fractioning of each term independently with a Gr\"obner basis containing only denominators that appear in that term, and once with a global Gr\"obner basis containing all $24$ denominators.
Note that these two approaches result in well defined but different output forms.
In both cases, we enable the option of our package to reduce the polynomials with {\form}, which was essential to achieve the best timing.
In Table~\ref{tab:timing} we list the resulting timings of these runs and compare features of the resulting output forms to those obtained with the algorithm of \cite{Boehm:2020ijp}.

The local Gr\"obner basis approach is most comparable to the algorithm used in Ref.\ \cite{Boehm:2020ijp}.
With this setup, our algorithm finished the partial fraction decomposition of all elements of the benchmark matrix in roughly $5.5$ hours on a single core of a desktop computer.
For the approach in which one calculates one global Gr\"obner basis for all $24$ denominators at once, our algorithm finds an optimized monomial ordering with which the Gr\"obner basis can be calculated within one minute in {\mathematica}. In this approach the reduction of all matrix elements took roughly $14.4$ hours on a Desktop computer, which is longer than in the local Gr\"obner basis approach.
We did not attempt to make a serious comparison of performance to the implementation of \cite{Boehm:2020ijp}, but the timing for this example seems to be of a similar order of magnitude compared to the runs of our package (Ref.\ \cite{Boehm:2020ijp} reported $27.9$ hours of runtime).

The resulting output forms are of similar overall size in all cases, see Table~\ref{tab:timing}.
Considering the rational functions as polynomials in $\mathbbm{Q}[q_1,\ldots,q_{24},c_2,\ldots,c_5,\epsilon]$, we determine the total degrees of the output forms in the different approaches.
In all cases, we consider fully expanded output forms and write them to disk with Mathematica.
As is shown in Table~\ref{tab:timing}, our method leads to lower degrees than the decomposition of Ref.\ \cite{Boehm:2020ijp}, as expected due to our choice of monomial ordering, cf.\ Section~\ref{sec:blockorder}.
For the comparison, it is noteworthy that our package does not treat non-linear denominators in a special way, they are fully taken into account in the decomposition.
Using a global Gr\"obner basis for the reduction yields a slightly larger result but has the big advantage that all terms are not only in a unique representation locally, but also globally.
Therefore sums of different coefficients which would appear in the calculation of the amplitude can be computed without any problems term by term, e.g.\ using {\form}.

\begin{table}[t]
    \centering
    \begin{tabular}{|l|c|c|c|c|}
    \hline
    Algorithm & Runtime & File size & Max.\ mon.\ deg.\ & Max.\ term length \\ 
    \hline
    Ref.\ \cite{Boehm:2020ijp} &   & $25.1$ MB & $20$ & 3564\\
    Global GB & $863$ min & $22.6$ MB & $12$ & 3109 \\
    Local GB & $356$ min & $21.3$ MB & $12$ & 3048 \\
    \hline
    \end{tabular}
    \caption{Comparison between our algorithm in the local and global reduction approach and the algorithm in Ref. \cite{Boehm:2020ijp}
    \label{tab:timing}}
\end{table}

%%%%%%%%%%%%%%%%%%%%%%%%%%%%%%%%%%%%%%%%%%%%%%%%%%%%%%%%%%%%%%%%%%%%%%%%%%%%%%%%%%%%%%%%
\section{Conclusion and outlook}
\label{sec:outlook}
%%%%%%%%%%%%%%%%%%%%%%%%%%%%%%%%%%%%%%%%%%%%%%%%%%%%%%%%%%%%%%%%%%%%%%%%%%%%%%%%%%%%%%%%

In this paper we introduced a method for the partial fraction decomposition of multivariate rational functions through polynomial reductions.
The method does not introduce spurious singularities and allows for a straightforward implementation in {\form} using local replacement rules.
We implemented our algorithms in a {\mathematica} package that is publicly available at \url{https://gitlab.msu.edu/vmante/multivariateapart}.

Our approach involves the construction of a special monomial ordering, which is optimized for computational performance and leads to results with comparably low degrees.
We showed for different ``real life'' examples how the algorithm helped to calculate Feynman amplitudes, in the presence of many kinematic variables, many denominators, or high polynomial degrees of the denominators.
In several of our examples, the calculation of the Gr\"obner basis is highly non-trivial and became possible through our method to define the monomial ordering.
The partial fractioned sum of terms lends itself to a straight-forward and compiler friendly implementation in numerical codes.

We discussed options for the reconstruction of rational functions from prime field samples.
Here, we proposed a new idea to reconstruct the denominator of a rational function based on knowledge about possible denominator factors and a small number of numerical samples.
This method has been successfully tested in the context of linear relations between Feynman integrals and reduced the number of samples required for the reconstruction of the full rational functions significantly.
Our partial fraction decomposition provides a systematic approach to the direct reconstruction rational function in this representation.
Future research might provide an optimized strategy to further reduce the number of samples required to reconstruct complicated rational functions in this approach.

Another future application of our methods could be the direct parametric integration of Feynman integrals~\cite{Brown:2008um,Panzer:2014caa}.
In this context it is important to recognize potential cancellations between different terms and to avoid the introduction of spurious (non-linear) denominators, since they can prevent a successful integration.
The algorithms discussed in this paper provide a systematic approach to this problem.

\section*{Acknowledgements}
We would like to thank Marco Besier, Bakul Agarwal and Federico Buccioni for helpful discussions and valuable feedback on applications of our package.
MH was supported in part by the German Research Foundation (DFG), through the Collaborative Research Center, Project ID 204404729, SFB 1044, and the Cluster of Excellence PRISMA$^+$, Project ID 39083149, EXC 2118/1.
AvM was supported in part by the National Science Foundation under Grants No.\ 1719863
and 2013859.

%%%%%%%%%%%%%%%%%%%%%%%%%%%%%%%%%%%%%%%%%%%%%%%%%%%%%%%%%%%%%%%%%%%%%%%%%%%%%%%%%%%%%%%%
\appendix
\addtocontents{toc}{\protect\contentsline{section}{Appendices}{}{}}

%%%%%%%%%%%%%%%%%%%%%%%%%%%%%%%%%%%%%%%%%%%%%%%%%%%%%%%%%%%%%%%%%%%%%%%%%%%%%%%%%%%%%%%%
\section{Polynomial reductions}
\label{sec:polyred}
%%%%%%%%%%%%%%%%%%%%%%%%%%%%%%%%%%%%%%%%%%%%%%%%%%%%%%%%%%%%%%%%%%%%%%%%%%%%%%%%%%%%%%%%

In this Appendix, we review some basic notions that can be found in any introductory textbook on the topic, see e.g.\ \cite{CoxLittleOShea,KreuzerRobbianoBook1}.

Let us consider polynomials in the variables $x_1,\ldots,x_n$ with coefficients in the field $K$ (e.g.\ the rational numbers).
These polynomials form a \emph{polynomial ring} $R=\mathbf{K}[x_1,\ldots,x_n]$.

A \emph{monomial ordering} is a total order $\prec$ on the set of monomials
\begin{equation}
    M \equiv \{x^\alpha \equiv x_1^{\alpha_1} \cdots x_n^{\alpha_n} | \alpha_i \in \mathbf{Z}_{\geq0} \}
\end{equation}
such that
\begin{itemize}
  \item[(i)] for all monomials $x^\alpha$, $x^\beta$ and $x^\gamma$:
$x^\alpha \leq x^\beta   \Rightarrow x^{\alpha+\gamma} \leq x^{\beta+\gamma}$,
  \item[(ii)] $1\leq x^\alpha\quad\forall\alpha$.
\end{itemize}

In many applications, in which the specific monomial ordering does not matter, one chooses the so-called \textit{degree reverse lexicographic order} (degrevlex), which first compares the total degree of two monomials, and then, if the degrees are the same, uses a lexicographic comparison and reverses the outcome of the latter.

There is a more general way of defining a monomial ordering, which will be relevant to us.
A monomial ordering is called a \emph{block ordering}, if one can group the variables into two sets
$\{x_1,\ldots,x_m\}$ and $\{y_{1},\ldots,y_n\}$, such that
\begin{equation}
x^\alpha \prec y^\beta\quad \forall\, \alpha, \beta~\text{with}~y^\beta \neq 1
\end{equation}
i.e. all monomials formed from variables from the first block are ``smaller'' than monomials formed from variables from the second block.

We call the ``greatest'' monomial of a polynomial $p$ with respect to a monomial order the \emph{leading monomial} of p.
A term is a monomial multiplied with a coefficient, which is a non-zero rational number for our purposes.
The \emph{leading term} $\lt(p)$ of polynomial $p$ is the term corresponding to the leading monomial.

Having defined a monomial ordering puts us in a position to discuss polynomial reductions.
Given two polynomials $p_1$ and $p_2$ together with a monomial ordering $\prec$, we
call $p_1$ reducible modulo $p_2$ if for some term $t$ of $p_1$ there is a term $u$ such that
$t=u\cdot \lt(p_2)$.
Then we say, \emph{$p_1$ reduces to $p_1'$ modulo $p_2$}, where
\begin{equation}
    p_1' = p_1 - u\cdot p_2
\end{equation}
The effect of the polynomial reduction is that a term in $p_1$ is replaced by a linear combination of ``smaller'' terms.

It is useful to introduce another concept.
A set of polynomials $g_1,...g_m \in R$, the \emph{generators}, define an \emph{ideal} $I \subset R$ as the set of all their linear combinations, where the coefficients are polynomials themselves,
$I = \{ \sum_{i} f_i g_i \text{~with~} f_i \in R \}$.
We also use the notation $I=\left< g_1,\ldots,g_m \right>$.
The choice of generators is not unique, and it is a non-trivial task to decide whether a polynomial is a member of a given ideal.
It is clear that, if a polynomial reduces to zero modulo the generators of an ideal that the polynomial is a member of that ideal.
The inverse, however, is not true and the remainder of a polynomial reduction will in general depend on the individual reduction steps.

It is possible to find a set of generators for an ideal, called a \emph{Gr\"obner basis}, which allows one to find a unique remainder for the reduction of any polynomial modulo the generators.
In this case, a reduction to zero occurs if and only if the polynomial is in the ideal.
A Gr\"obner basis can be calculated algorithmically by adjoining specific differences of generators.
The Gr\"obner basis depends on the choice of the monomial ordering and is unique if reduced with respect to itself.

%%%%%%%%%%%%%%%%%%%%%%%%%%%%%%%%%%%%%%%%%%%%%%%%%%%%%%%%%%%%%%%%%%%%%%%%%%%%%%%%%%%%%%%%%
\section{\texorpdfstring{{\Leinartas}'}{Leinartas'} requirements and polynomial reductions}
\label{sec:leinartaspolyred}
%%%%%%%%%%%%%%%%%%%%%%%%%%%%%%%%%%%%%%%%%%%%%%%%%%%%%%%%%%%%%%%%%%%%%%%%%%%%%%%%%%%%%%%%%

In this Section we revisit the two central decomposition steps in Algorithm~\ref{alg:leinartas} leading to {\Leinartas}' decomposition, see Section \ref{sec:leinartas} and consider under which circumstances they are reproduced by the polynomial reductions in Algorithm~\ref{alg:mapart}, see Section \ref{sec:alg}.
As we will show, Algorithm~\ref{alg:mapart} may or may not produce a {\Leinartas}' decomposition, depending on the monomial ordering.
If the monomial ordering sorts first for the $q_i$ and then for the $x_i$, requirement (i) will be filfilled, that is, the denominator zeros will be separated.
If the $q_i$ are sorted lexicographically, also requirement (ii), the algebraic independence of different denominator factors, will be guaranteed~\cite{Abreu:2019odu}.
For degree based orderings of the $q_i$, requirement (ii) is violated in general.
This means in particular, that our monomial block ordering proposed in Section \ref{sec:blockorder} guarantees only requirement (i) and not (ii), and will therefore not lead to a {\Leinartas} decomposition.
We emphasize, that an even more general choice of monomial ordering will ensure neither of the two requirements, but still allows for a unique output form.

First, we consider the requirement (i), i.e.\ that the denominators of each term shall have common zeros in $\overline{K}^n$.
Let us assume we encounter a term which is not fully decomposed yet and has denominators $\{d_1(x_1,\ldots),\ldots,d_m(x_1,\ldots)\}$.
By Hilbert's Nullstellensatz, a finite set of polynomials $\{d_1^{\alpha_1},\ldots,d_m^{\alpha_m}\}$ has no common zeros in $\overline{K}^n$ if and only if there exist polynomials $h_1,\ldots,h_m$, such that
\begin{equation}
1 = \sum_i h_i(x_1,\ldots)\, d^{\alpha_i}_i(x_1,\ldots).
\label{eq:hilbert_init}
\end{equation}
Dividing this equation by $d_1^{\alpha_1}\cdots d_m^{\alpha_m}$ gives the decomposition \cite{Raichev:2012pf}
\begin{equation}
    \frac{1}{d_1^{\alpha_1}\cdots d_m^{\alpha_m}} = \sum_i \frac{h_i(x_1,\ldots)} {d_1^{\alpha_1} \cdots \hat{d_i}^{\alpha_i} \cdots d_m^{\alpha_m}}.
\end{equation}
where $\hat{d_i}^{\alpha_i}$ means leaving $d_i^{\alpha_i}$ out in the product.

This decomposition step in Algorithm~\ref{alg:leinartas} has a direct analogue in terms of a polynomial reduction in step 5 of Algorithm~\ref{alg:mapart} for a suitable monomial ordering.
Multiplying Eq.\ \eqref{eq:hilbert_init} by $q_1^{\alpha_1}\cdots q_m^{\alpha_1}$ and replacing $q_i d_i(x_1,\ldots) = 1$ gives
\begin{equation}
    q_1^{\alpha_1}\cdots q_m^{\alpha_m} - \sum_i h_i(x_1,\ldots)\, q_1^{\alpha_1}\cdots\hat{q_i}^{\alpha_i}\cdots q_m^{\alpha_m} = 0,
\label{eq:hilbert_red}
\end{equation}
where $\hat{q_i}^{\alpha_i}$ means leaving $q_i^{\alpha_i}$ out in the product.
It is clear that the polynomial on the lhs of Eq.\ \eqref{eq:hilbert_red} is a member of the ideal $I=\left<1-q_1 d_1(x_1,\ldots), \ldots\right>$.
Furthermore, $q_1^{\alpha_1}\cdots q_m^{\alpha_m}$ is the leading monomial for any monomial ordering which sorts first for all of the $q_1,\ldots,q_m$ and then for the $x_1,\ldots,x_n$.
In Algorithm~\ref{alg:mapart}, that leading monomial will be thus be replaced by the subleading terms on the lhs of Eq.\ \eqref{eq:hilbert_red}, resulting in terms with fewer factors of $q_i$, that is, simpler denominators.
A fully reduced term will therefore not have all factors $q_1\cdots q_n$.
This proves the above statements 
We conclude that, for our choice of monomial ordering, the polynomial reduction in Algorithm~\ref{alg:mapart} guarantees requirement (i) of {\Leinartas}' decomposition, that is, the separation of independent denominator zeros.

Second, we consider requirement (ii), that is, the algebraic independence of the denominator factors.
Let us assume that our decomposition contains a term with denominator $d_1^{\alpha_1}\cdots d_m^{\alpha_m}$, where the factors $\{d_1,...d_m\}$ are algebraically dependent.
One can show that then also the polynomials $\{d_1^{\alpha_1},...d_m^{\alpha_m}\}$ are algebraically dependent and there exists a non-zero annihilating polynomial $p(y_1,\ldots,y_m)$ with
\begin{equation}
p(d_1^{\alpha_1},\ldots,d_m^{\alpha_m})=0  
\label{eq:algdep}
\end{equation}
after inserting the explicit expressions the $d_i(x_1,\ldots)$.
In Algorithm~\ref{alg:leinartas}, one can for example choose one of the monomials of $p$ with smallest degree and substitute the remainder, i.e.\ one has
\begin{align}
c_\beta(d^\alpha)^\beta &= - \sum_{\gamma\in S} c_\gamma (d^\alpha)^\gamma,
\label{eq:alg_dec_0}
\end{align}
where $d=(d_1,...d_m)$, $\alpha$, $\beta$ and $\gamma$ are multi-indices in $\mathbbm{N}^m$, with $\sum_i \beta_i \leq \sum_i \gamma_i$ for all $\beta$ in the set of multi-indices $S$.
Dividing by $c_\beta d^{\beta+1}$ gives the decomposition step~ \cite{Raichev:2012pf}
\begin{equation}
    \frac{1}{d_1^{\alpha_1}\cdots d_m^{\alpha_m}}
    = -\sum_{\gamma\in S} \frac{c_\gamma}{c_\beta}
    \prod_{i=1}^m\frac{d_i^{\alpha_i\gamma_i}}{d_i^{\alpha_i(\beta_i+1)}}.
\label{eq:alg_dec_1}
\end{equation}
Since $\sum_i \beta_i \leq \sum_i \gamma_i$ and $\beta\neq\gamma$, for each $\gamma$ there exists an $i$ such that $\beta_i+1\leq \gamma_i$ and the factor $d_i^{\beta_i+1}$ is removed from the denominator.
Therefore, each term on the rhs of \eqref{eq:alg_dec_1} depends on at least one denominator factor $d_i$ less than the lhs.

The described decomposition step in Algorithm~\ref{alg:leinartas} may or may not have an analogue in terms of a polynomial reduction in step 5 of Algorithm~\ref{alg:mapart}, depending on the monomial ordering.
In general, for a degree based ordering of the $q_i$ variables, either globally or in blocks like proposed in this paper, a similar decomposition will \emph{not} occur.
In \eqref{eq:alg_dec_1}, the denominator degree may actually be higher on the rhs than on the lhs, and it may therefore not correspond to a polynomial reduction for such an ordering.
A lexicographic ordering, on the other hand, \emph{will} separate algebraically dependent denominators.
To show this, we pick in Eq.\ \eqref{eq:algdep} the unique exponent $\beta'$ of $p$ such that $(q^\alpha)^{\beta'}$ is minimal with respect to our monomial ordering.
Multiplication with $(q^\alpha)^{\beta'+1}/(c_{\beta'})$ gives 
\begin{equation}
    q_1^{\alpha_1}\cdots q_m^{\alpha_m}
    + \sum_{\gamma\in S} \frac{c_\gamma}{c_{\beta'}} \prod_{i=1}^m {d_i^{\alpha_i\gamma_i}}{q_i^{\alpha_i(\beta'_i+1)}} = 0.
\end{equation}
Replacing all products $q_i d_i(x_1,\ldots) = 1$ gives
\begin{equation}
    q_1^{\alpha_1}\cdots q_m^{\alpha_m}
    +\sum_{\gamma\in S} \frac{c_\gamma}{c_{\beta'}} \prod_{i=1}^m
    d_i^{\max(\alpha_i (\gamma_i - \beta'_i - 1), 0)} q_i^{\max(\alpha_i (\beta'_i + 1 - \gamma_i), 0)}  = 0.
\label{eq:alg_red}
\end{equation}
Assuming a lexicographic ordering of the $q_i$ with $q_1 \succ q_2 \succ \ldots$ implies that for each term $\gamma$ there exists a $j=1,\ldots,m$ such that $\beta'_i = \gamma_i$ for all $i=1,\ldots,j$, $\beta'_j \prec \gamma_j$, the powers of $q_i$ coincide with that of the first term for $i=1,\ldots,j-1$, but $q_j$ is removed.
This means that the first term in \eqref{eq:alg_red} is indeed the leading term.
Since furthermore the polynomial on the lhs of \eqref{eq:alg_red} is a member of the ideal $I=\left<q_1 d_1(x_1,\ldots)-1, \ldots \right>$, the first term in \eqref{eq:alg_red} would be reduced in step 5 of Algorithm~\ref{alg:mapart} for this ordering.
We conclude that, for the block monomial ordering presented in this article, the polynomial reduction in Algorithm~\ref{alg:mapart} does not guarantee requirement (ii) of {\Leinartas}' decomposition, that is, algebraically independent denominators.
In contrast, a lexicographical ordering guarantees this requirement.

We would like to illustrate the impact of the monomial ordering on the decomposition with the following example. We consider the irreducible denominator factors
\begin{equation}
    d_1 = x^3+y^4,\quad d_2 = x+y^2,\quad d_3 = x^2+y,
\label{eq:algdepdi}
\end{equation}
which share a common zero.
We note that three denominator factors in two variables must be algebraically dependent on general grounds.
Let us consider the ideal
\begin{equation}
    I=\left<q_1 d_1(x,y)-1, q_2 d_2(x,y)-1, q_3 d_3(x,y)-1\right>.
\end{equation}
The monomial block ordering proposed in this article gives
\begin{equation}
    \{\{q_1, q_2, q_3\}, \{x, y\} \} .
\end{equation}
Calculating the Gr\"obner basis of $I$ with respect to this ordering, we see that the polynomial representation of $1/(d_1 d_2 d_3)$,
\begin{equation}
q_1 q_2 q_3,
\label{eq:q1q2q3}
\end{equation}
is fully reduced already, despite the denominators $d_1$, $d_2$ and $d_3$ being algebraically dependent.

Next, let us consider a lexicographic ordering of the $q_i$ according to
\begin{equation}
    \{ \{q_1\}, \{q_2\}, \{q_3\}, \{x, y\} \}.
\end{equation}
In this case, the Gr\"obner basis computation of $I$ reveals that $q_1 q_2 q_3$ is reducible.
Alternatively, we can derive a reduction identity of type \eqref{eq:alg_red} as described above.
Indeed, it is not difficult to calculate the annihilator
\begin{align}
&p(y_1, y_2, y_3) = y_1^4 - 4 y_1^3 y_2^2 - 4 y_1^3 y_3 + 3 y_1^3 + 6 y_1^2 y_2^4 +  4 y_1^2 y_2^2 y_3 - y_1^2 y_2^2 + 8 y_1^2 y_2 y_3^2
\notag\\ &\quad
- 6 y_1^2 y_2 y_3 - 
 8 y_1^2 y_2 - 2 y_1^2 y_3^3 + 6 y_1^2 y_3^2 - 2 y_1^2 y_3 + 3 y_1^2 - 
 4 y_1 y_2^6
 + 4 y_1 y_2^4 y_3
\notag\\ &\quad
 - 7 y_1 y_2^4 - 16 y_1 y_2^3 y_3^2 + 
 12 y_1 y_2^3 y_3 - 8 y_1 y_2^3 + 4 y_1 y_2^2 y_3^3 + 4 y_1 y_2^2 y_3^2 + 
 36 y_1 y_2^2 y_3
\notag\\ &\quad
 - 16 y_1 y_2 y_3^3 - 18 y_1 y_2 y_3^2 + 2 y_1 y_2 y_3 - 
 5 y_1 y_2 + 4 y_1 y_3^4 + 2 y_1 y_3^3 - y_1 y_3^2 + 2 y_1 y_3 
\notag\\ &\quad
+ y_2^8 - 
 4 y_2^6 y_3 + 5 y_2^6 + 8 y_2^5 y_3^2 - 6 y_2^5 y_3 - 2 y_2^4 y_3^3 + 
 6 y_2^4 y_3^2 - 2 y_2^4 y_3 + 5 y_2^4 
\notag\\ &\quad
- 16 y_2^3 y_3^3 - 14 y_2^3 y_3^2 - 
 2 y_2^3 y_3 + 20 y_2^2 y_3^4 + 6 y_2^2 y_3^3 - 8 y_2 y_3^5 + 5 y_2 y_3^4 + 
 y_3^6 - 2 y_3^5
% TeXed terms are ordered lexicographically
\end{align}
with $p(d_1(x,y),d_2(x,y),d_3(x,y))=0$ and identify $q_i d_i = 1$ modulo $I$ to obtain
\begin{align}
&q_1 q_2 q_3 
-\tfrac{1}{2}q_1 q_2
- \tfrac{1}{2}(d_2^7 + 5 d_2^5 + 5 d_2^3)  q_1 q_3^6
+ (2 d_2^5 + 3 d_2^4 + d_2^3 + d_2^2) q_1 q_3^5
 \notag\\ &
- (4 d_2^4 + 3 d_2^3 - 7 d_2^2) q_1 q_3^4
+ (d_2^3 + 8 d_2^2 - 3 d_2) q_1 q_3^3
- \tfrac{1}{2}(20 d_2 + 5) q_1 q_3^2
+ 4 q_1 q_3
 \notag\\ &
- \tfrac{1}{2}(d_1^3 + 3 d_1^2 + 3 d_1) q_2 q_3^6
+ (2 d_1^2 + d_1 - 1) q_2 q_3^5
- \tfrac{1}{2}(6 d_1 - 1) q_2 q_3^4
+ (d_1 - 1) q_2 q_3^3
 \notag\\ &
- 2 q_2 q_3^2
+ \tfrac{1}{2}(4 d_1^2 d_2 -6 d_1 d_2^3 + d_1 d_2 + 8 d_1 + 4 d_2^5 + 7 d_2^3 + 8 d_2^2 + 5) q_3^6
 \notag\\ &
- (2 d_1 d_2 - 3 d_1 + 2 d_2^3 + 6 d_2^2 + 18 d_2 + 1 ) q_3^5
- (4 d_1 - 8 d_2^2 + 2 d_2 - 9) q_3^4
 \notag\\ &
- (2 d_2 - 8) q_3^3
=0,
 \end{align}
where $d_1$, $d_2$, $d_3$ are meant to be replaced by their definitions \eqref{eq:algdepdi}.
As we see, this gives a reduction identity for \eqref{eq:q1q2q3} since the first term is indeed the leading term in this ordering, such that the algebraically dependent denominators are decomposed.
However, we also see that this decomposition leads to a significant increase in the degrees of the polynomials.

\bibliographystyle{JHEP}
\bibliography{partialfractions}

\end{document}